\documentclass{article}

\usepackage{times}
\usepackage{color}
\usepackage{multicol}
\usepackage{pdfsync}
\usepackage{pgf}
\usepackage{tikz}
\usepackage{natbib}
\usepackage{amssymb}
\usepackage{graphicx}
\usepackage{color}
\usepackage{hhline}
\usepackage{amsmath}
\usepackage{proceed2e}
\usepackage{enumerate}
\newcommand{\ignore}[1]{}

\newcommand{\boxtheorem}{\hfill $\Box$\vspace{3mm}}
\newcommand{\nit}[1]{{\it #1}}

\usepackage{graphicx}

\usepackage{algpseudocode}
\usepackage{algorithm}
\usetikzlibrary{topaths,calc}
\newcommand{\red}[1]{\textcolor{red}{#1}}

\newcommand{\combabak}[1]{{\vspace{4mm}\noindent \bf  COMM(Babak):}~ \red {\em  #1}\hfill {\bf END.}\\}

\ignore{  }

\ignore{
\newcounter{lemmaA-counter}
\newcounter{propositionA-counter}
\setcounter{lemmaA-counter}{0}

\setcounter{propositionA-counter}{0}

{\vskip \abovedisplayskip \refstepcounter{lemmaA-counter}%
\noindent {\bf Lemma A.\arabic{lemmaA-counter}.}}%

{\vskip \abovedisplayskip \refstepcounter{propositionA-counter}%
\noindent {\bf Proposition A.\arabic{propositionA-counter}.}}%
}

\newcommand{\defproof}[2]{{\noindent\bf Proof of #1:\
}#2 \boxtheorem}

\makeatother

\newcommand{\mc}[1]{\mathcal{ #1}}

\usepackage{amsmath,amsfonts,amssymb,amsthm,epsfig,epstopdf,titling,url,array}

\newcommand{\sm}[1]{{\small \text{\nit{#1}}}}
\theoremstyle{plain}
\newtheorem{theorem}{Theorem}[section]

\newtheorem{proposition}[theorem]{Proposition}

\theoremstyle{definition}
\newtheorem{definition}{Definition}[section]
\newtheorem{example}{Example}[section]

\theoremstyle{remark}

\usepackage{times}

\title{\bf Query-Answer Causality in Databases: Abductive Diagnosis and View-Updates}%\thanks{\ignore{To appear in Proc. UAI Causal Inference Workshop, 2015. }{\small \tt \{bsalimi,bertossi\}@scs.carleton.ca}}}

\date{}

\author{ {\bf Babak Salimi} \ and \ {\bf Leopoldo Bertossi} \\
School of Computer Science \\
Carleton University\\
Ottawa, Canada.\\
{\small \tt \{bsalimi,bertossi\}@scs.carleton.ca}
}
\begin{document}

\maketitle
\thispagestyle{empty}
\pagestyle{plain}

\begin{abstract} \vspace{-5mm}
Causality has been recently introduced in databases, to model, characterize and possibly compute causes
for query results (answers). Connections between query causality and consistency-based diagnosis and database repairs
(wrt. integrity constrain violations) have been established in the literature. In this work we establish connections between
query causality and abductive diagnosis and the view-update problem. The unveiled relationships allow us to obtain
new complexity results for query causality -the main focus of our work- and also for the two other areas.
\end{abstract}

Causality is an important notion that appears at the foundations of many scientific disciplines, in the practice of technology, and also in our everyday life. Causality
is unavoidable to understand and manage {\em uncertainty} in data, information, knowledge, and theories.  In data management in particular, there is a need to represent, characterize
 and compute the causes that explain why certain query results are obtained or not, or why natural semantic conditions, such as integrity constraints, are not satisfied. Causality can also
 be used to explain the contents of a view, i.e. of a predicate with virtual contents that is defined in terms of other physical, materialized relations (tables).

 In this
 work we concentrate on causality as defined for- and applied to relational databases. Most of the work on causality has been developed in the context of knowledge representation,
 and little has been said about causality in data management. Furthermore, in a world of big uncertain data, the necessity to understand the data beyond simple query answering, introducing explanations in different  forms, has become particularly relevant.

The notion of causality-based explanation for a query result  was introduced in \citep{Meliou2010a}, on the basis of the deeper concept of {\em actual causation}.\footnote{In contrast with general causal claims, such as ``smoking causes cancer", which refer some sort of related events,  actual causation  specifies a particular instantiation of a causal relationship,  e.g., ``Joe's smoking is a  cause for his cancer".}
Intuitively, a tuple (of constants) $t$  is an {\em actual cause} for an answer $\bar{a}$ to a
conjunctive query $\mc{Q}$ from  a relational database instance $D$ if there is a ``contingent" subset of tuples $\Gamma$, accompanying $t$,
such that, after removing $\Gamma$ from $D$, removing $t$ from $D\smallsetminus \Gamma$ causes $\bar{a}$ to switch from being an answer to being a non-answer (i.e.
not being an answer). Usually, actual causes and contingent tuples  are restricted to be among a pre-specified set
of {\em endogenous tuples}, which are admissible, possible candidates for causes, as opposed to {\em exogenous tuples}.

A cause $t$ may have different associated contingency sets $\Gamma$. Intuitively, the smaller they are the strongest is $t$ as a cause (it need less company to
undermine the query answer). So, some causes may be stronger than others. This idea is formally captured through the notion of {\em causal responsibility},
and introduced in \citep{Meliou2010a}. It reflects the relative degree of actual causality. In applications involving large
data sets, it is crucial to rank potential causes according to their responsibilities \citep{Meliou2010b,Meliou2010a}.

Furthermore,  {\em view-conditioned causality} was proposed in \citep{Meliou2010b,Meliou2011} as a restricted form of query causality, to determine causes for a set of unexpected
query results, but  conditioned to the correctness of prior knowledge about some other set of results.

Actual causation, as used in  \citep{Meliou2010a,Meliou2010b,Meliou2011}, can be traced back to
\citep{Halpern01,Halpern05}, which provides  a model-based account of causation on the basis of {\em counterfactual dependence}.\footnote{As discussed in \citep{icdt15}, some objections
to the Halpern-Pearl model of causality and the corresponding changes \citep{halpern14,halpern15} do not affect results in the context of databases.} {\em Causal responsibility} was introduced  in \cite{chocker04}, to provide a graded,  quantitative notion of causality when multiple causes may over-determine an outcome.

\ignore{
Apart from the explicit use of causality, research on explanations for query results has focused mainly, and rather implicitly, on provenance
\citep{BunemanKT02,Cheney09,tannen}, and
more recently, on provenance for non-answers \citep{ChapmanJ09,HuangCDN08}.
}
\ignore{
For connections between
causality and provenance, see \citep{Meliou2010a,Meliou2010b}. However, causality is a more refined notion that identifies causes
for query results on the basis of  user-defined criteria, and ranks causes according to
their responsibility \citep{Meliou2010b}.
}

Model-based diagnosis \cite[sec. 10.3]{struss}, an area of knowledge representation, addresses the problem of, given the {\em specification} of a system in some logical formalism and a usually unexpected {\em observation}  about the system, obtaining {\em explanations} for the observation, in the form of a diagnosis for the unintended behavior. Since this and
causality are related to explanations,  a first connection between causality and {\em consistency-based diagnosis} \citep{Reiter87}, a form of model-based diagnosis, was established
in  \citep{nmr14,icdt15}: Causality  and the responsibility problem can be formulated as {\em consistency-based diagnosis} problems, which allowed to extend the results
in  \citep{Meliou2010a}. However, no precise connection has been established so far between causality and {\em abductive diagnosis} \citep{console91,EiterGL95}, another form of model-based diagnosis.

The definition of causality for query answers applies to monotone queries \citep{Meliou2010a,Meliou2010b}. However, all complexity and algorithmic results in \citep{Meliou2010a,icdt15} have
been  restricted to first-order (FO) monotone queries. Other important classes of monotone queries, such as Datalog queries \citep{ceri90,Abiteboul95}, possibly with recursion, require further investigation.

In \citep{icdt15} connections were established between query causality, database repairs \citep{2011Bertossi}, and consistency-based diagnosis. In particular, complexity results for several causality
problems were obtained from the repair connection. In the line of this kind of research, in this work we unveil  natural connections between actual causation and {\em abductive diagnosis},
and also the view-update problem in databases (more on this latter connection later in the section).

As opposed to consistency-based diagnoses, which is  usually practiced with  FO specifications, abductive diagnosis is commonly performed under a logic programming (LP) approach (in the general sense of LP) to knowledge representation \citep{DeneckerK02,EiterGL97,Gottlob10b}. Since Datalog can be seen as a form of LP, we manage to extend and formulate  the notion of  query-answer causality to  Datalog queries via the
  abductive diagnosis connection, in this way extending causality to a new class of queries, e.g. recursive queries, and obtaining complexity results on causality for them.

  Abductive reasoning/diagnosis has been applied to the view update problem in databases \citep{Kakas90,Console95}, which is about characterizing and computing updates of physical
  database relations that give an account of (or have as result) the intended updates on views. The idea is that abductive diagnosis  provides (abduces) the reasons for the desired view updates,
  and they are given as changes on base tables.

  In this work we also explore fruitful connections of causality with this  {\em view-update problem} \citep{Abiteboul95}, i.e.
   about updating a database through views. An important aspect of the problem is that one wants  the base, source database, i.e. the base relations, to change in a minimally way
   while still producing the view updates.
    Put in different terms, it is an update propagation problem, from
views to base relations. This classical and important problem in databases.

The {\em delete-propagation} problem  \citep{BunemanKT02,Kimelfeld12a,Kimelfeld12b} is a particular case of the view-update problem where only tuple deletions are allowed
on/from the views. If the views are defined by monotone queries, only database deletions can give an account of view deletions. So, in this case, a minimal set (in some sense) of deletions
from the base relations is expected to be performed. This is ``minimal source-side-effect" case. It is also possible to consider minimizing the side-effect on the
 view, which also requires that other tuples in the (virtual) view contents are not affected (deleted)  \citep{BunemanKT02}.

 In this work we provide a precise connection between different variants of the delete-propagation problem and query causality. In particular, we show that the minimal source-side-effect problem is related to the {\em most-responsible cause problem}, which was formulated and investigated in \citep{icdt15}; and also  that the ``minimal view side-effect problem"
 is related to view-conditioned causality we already mentioned above.

The established connections between abductive diagnoses, query causality and delete-propagation problems allow us to adopt (and possibly adapt) established
 results for some of them for application to the others. In this way we obtain some new complexity results.

More precisely, our main results  are as follows:\footnote{The possible connections between the areas and problems in this paper were suggested
 in \citep{buda14}, but no precise results were formulated there.}

\vspace{-3mm}
 \begin{enumerate}
 \item
  We establish precise connections between causality for Datalog queries and abductive diagnosis.
 More precisely, we establish mutual characterizations of each in terms of the other, and computational reductions,
 between actual causes for Datalog queries and abductive diagnosis from Datalog specifications.

 We profit from these connections to obtain new algorithmic and complexity results for each of the two problems separately.

 \begin{enumerate}
 \item
 We characterize and obtain causes in terms of- and from abductive diagnoses.

 \item \ignore{ The connection between causality and abduction allows us to apply results and techniques from each to the other.}
  We show that deciding tuple causality
for Datalog queries, possibly recursive, is \nit{NP}-complete in data.

 \item We identify a class of Datalog queries for which deciding causality is tractable in combined complexity.

\end{enumerate}

 \item We establish and profit from precise connections between delete-propagation and causality. More precisely, we show that:
 \begin{enumerate}
 \item Most-responsible causes and  view-conditioned  causes can be obtained from solutions to different variants of
 the delete-propagation problem and vice-versa.

 \item Computing the size of the solution to a minimum source-side-effect problem is hard for $FP^\nit{NP(log(n))}$.

 \item Deciding weather an answer has a view-conditioned cause is \nit{NP}-complete.

  \item We can identify some new classes of queries for which computing minimum source-side-effect delete-propagation is tractable.
  \end{enumerate}
\end{enumerate}

\section{PRELIMINARIES AND CAUSALITY DECISION PROBLEMS}\label{sec:prel}

\vspace{-2mm}
We  consider relational database schemas of the form $\mathcal{S} = (U,\mc{P})$,  where $U$ is the possibly infinite
database domain and $\mc{P}$ is a finite set of {\em database predicates}\footnote{As opposed to built-in predicates (e.g. $\neq$) that we assume
do not appear, unless explicitly stated otherwise.} of fixed arities. A database instance $D$
compatible with $\mathcal{S}$ can be seen as a finite set of ground atomic formulas (in databases aka. atoms or tuples), of the form $P(c_1, ..., c_n)$, where $P \in \mc{P}$ has arity $n$, and the constants $c_1, \ldots , c_n \in U$.

A {\em conjunctive query}  (CQ) is a formula $\mc{Q}(\bar{x})$ of the first-order (FO) language $\mc{L}(\mc{S})$ associated to $\mc{S}$ of the form \ $\exists \bar{y}(P_1(\bar{s}_1) \wedge \cdots \wedge P_m(\bar{s}_m))$,
where the $P_i(\bar{s}_i)$ are atomic formulas, i.e. $P_i \in \mc{P}$, and the $\bar{s}_i$ are sequences of terms, i.e. variables or constants of $U$. The $\bar{x}$ in  $\mc{Q}(\bar{x})$ shows
all the free variables in the formula, i.e. those not appearing in $\bar{y}$.  A sequence $\bar{c}$ of constants is an answer to query $\mc{Q}(\bar{x})$ if $D \models \mc{Q}[\bar{c}]$, i.e.
the query becomes true in $D$ when the variables are replaced by the corresponding constants in $\bar{c}$. We denote the set of all answers to an open conjunctive query $\mc{Q}(\bar{x})$ with $\mc{Q}(D)$.

A conjunctive query is {\em boolean} (a BCQ), if $\bar{x}$ is empty, i.e. the query is a sentence, in which case, it is true or false
in $D$, denoted by $D \models \mc{Q}$ and $D \not\models \mc{Q}$, respectively.
When $\mc{Q}$ is a BCQ, or contains no free variables, $\mc{Q}(D) = \{\nit{yes}\}$ if $\mc{Q}$ is true, and $\mc{Q}(D) = \emptyset$, otherwise.

A query $\mc{Q}$ is {\em monotone} if for every two instances $D_1 \subseteq D_2$, \ $\mc{Q}(D_1) \subseteq \mc{Q}(D_2)$, i.e. the set of answers grows monotonically with the instance. For example, CQs and unions of CQ (UCQs) are monotone queries. Datalog
queries \citep{ceri90,Abiteboul95}, although not FO, are also monotone (cf. Section \ref{sec:causeintro} for more details).

\vspace{-2mm}
\subsection{CAUSALITY AND RESPONSIBILITY} \label{sec:causeintro}

\vspace{-2mm}

In the rest of this work, unless otherwise stated, we will assume that a database instance $D$ is split in two disjoint sets, $D=D^n \cup D^x$,  where $D^n$ and $D^x$ denote the sets of {\em endogenous} and {\em exogenous} tuples,
 respectively;  and $\mc{Q}$ is a monotone query. %  with $a \in \mc{Q}(D)$, causes for the answer $a$ is defined in \citep{Meliou2010b} as follows:

 \begin{definition}  \label{def:querycause}   A tuple $\tau \in D^n$ is  a
{\em counterfactual cause} for an answer $\bar{a}$ to $\mc{Q}$ in $D$  if $D\models \mc{Q}(\bar{a})$ and $D\smallsetminus \{\tau\}  \not \models \mc{Q}(\bar{a})$.  A tuple $\tau \in D^n$ is an {\em actual cause} for  $\bar{a}$
if there  exists $\Gamma \subseteq D^n$, called a {\em contingency set},  such that $\tau$ is a counterfactual cause for $\bar{a}$ in $D\smallsetminus \Gamma$.  \boxtheorem
\end{definition}

\vspace{-5mm}
{\em $\nit{Causes}(D,\mc{Q}(\bar{a}))$
denotes the set of actual causes for $\bar{a}$}.  This set  is non-empty on the assumption that $\mc{Q}(\bar{a})$ is true in $D$. When the query $\mc{Q}$ is boolean, $\nit{Causes}(D,\mc{Q})$ contains the causes for the answer $\nit{yes}$ in $D$.

The definition of query-answer causality can be applied without any conceptual changes to Datalog queries. In the case of a Datalog, the query $\mc{Q}(\bar{x})$ is a whole program $\Pi$ that accesses an underlying extensional database $E$ that is not part of the query. Program $\Pi$ contains a rule that defines
a top answer-collecting predicate $\nit{Ans}(\bar{x})$. Now,  $\bar{a}$ is an answer to query $\Pi$ on $E$ when $\Pi \cup E \models \nit{Ans}(\bar{a})$. Here, entailment ($\models$) means that the RHS belongs to the minimal model of the LHS.
A Datalog query is boolean if the top answer-predicate is propositional, say $\nit{ans}$. In the case of Datalog, we sometimes use the notation $\nit{Causes}(E,\Pi(\bar{a}))$ or $\nit{Causes}(E,\Pi)$, depending on whether $\Pi$ has a $\nit{Ans}(\bar{x})$ or $\nit{ans}$ as answer predicate, resp.

Given a $\tau \in \nit{Causes}(D,\mc{Q}(\bar{a}))$, we collect all
subset-minimal  contingency sets associated with $\tau$:
{\small
\begin{eqnarray}
\nit{Cont}(D, \mc{Q}(\bar{a}),\tau)\!\!\!&:=&\!\!\!\{  \Lambda\subseteq D^n~|~ D\smallsetminus \Lambda \nonumber
\models Q(\bar{a}),\\ \nonumber
&&~D\smallsetminus (\Lambda \cup \{\tau\}) \not \models \mc{Q}(\bar{a}),   \mbox{ and }
\label{eq:ct}\\
   &&\!\!\!\forall \Lambda'\subsetneqq \Lambda, \ D \smallsetminus (\Lambda' \cup \{\tau\})
\models \mc{Q}(\bar{a}) \}. \nonumber
\end{eqnarray}
} The {\em responsibility} of actual cause $\tau$ for answer $\bar{a}$, denoted  $\rho_{_{\!\mc{Q}(\bar{a})\!}}(\tau)$,  is $\frac{1}{(|\Gamma| + 1)}$, where $|\Gamma|$ is the
size of the smallest contingency set for $\tau$. Responsibility can be extend to all tuples in $D^n$ by setting their value to $0$, and they are not actual causes for $\mc{Q}$.

\vspace*{1mm}
\begin{example}\label{ex:cfex1}
Consider a database $D$ with  relations \sm{Author}(\sm{Name,Journal}) and \sm{Journal}(\sm{JName},\sm{Topic,\#Paper}), and contents as
below:

{\tiny
 \begin{tabular}{l|c|c|} \hline
Author & \nit{Name} & \nit{JName} \\\hline
 & Joe  & TKDE\\
& John & TKDE\\
& Tom & TKDE\\
& John & TODS\\
 \hhline{~--} \end{tabular} ~~~~~~~\begin{tabular}{l|c|c|c|} \hline
Journal  & \nit{JName} & \nit{Topic}& \nit{\#Paper} \\\hline
 & TKDE & XML & 30\\
& TKDE & CUBE& 31\\
& TODS & XML & 32\\
 \hhline{~---}
\end{tabular}
} \\ \\
Consider the conjunctive query:

\vspace{-5mm}
{\small
\begin{eqnarray}
\mc{Q}(\sm{Name}, \sm{Topic})\!:&&\!\!\!\!\!\!\!\!\!\!\!\! \exists \sm{Journal~JName~\#Paper}(\sm{Author}(\sm{Name,JName}) \nonumber \\
&&\hspace*{0.5cm}\ \wedge \ \sm{Journal}(\sm{JName,Topic,\#Paper}), \label{eq:query}
\end{eqnarray}
}
\vspace{-10mm}

\phantom{po}

\noindent which has the following answers:

\vspace{-5mm}
{\tiny
\hspace{4.8cm} \begin{tabular}{l|c|c|} \hline
$\mc{Q}(D)$ & \nit{Name} & \nit{Topic} \\\hline
 & Joe  & XML\\
  & Joe  & CUBE\\
& Tom & XML\\
& Tom & CUBE\\
& John & XML\\
& John & CUBE\\
 \hhline{~--}
 \end{tabular} }
%\end{center}

Assume   $\langle \sm{John, XML} \rangle$ is an unexpected answer to $\mc{Q}$, and we want to compute its causes assuming that all tuples are endogenous.

It turns out that
 \sm{Author(John, TODS)} is an actual cause, with contingency sets  $\Gamma_1 = \{\sm{Author(John, TKDE)}\}$ and $\Gamma_2$=\{\sm{Journal(TKDE, XML, 32)}\}, because \linebreak \sm{Author(John, TODS)}  is a counterfactual cause for answer $\langle \sm{ John, XML}\rangle$ in both of $D \smallsetminus \Gamma_1$ and $D \smallsetminus \Gamma_2$.  Therefore, the responsibility of \sm{Author(John, TODS)} is $\frac{1}{2}$.

Likewise, \sm{ Journal(TKDE, XML, 32)},  \sm{Author(John, TKDE)}, \sm{Journal(TODS,XML, 32)}  are actual causes  for $\langle \sm{John, XML} \rangle$ with responsibility $\frac{1}{2}$.

Now, under the assumption that the tuples in $\sm{Journal}$ are the endogenous tuples,
the only actual causes for answer $\langle \sm{John, XML} \rangle$  are \sm{ Author(John, TKDE)}
and \sm{Author(John, TODS)}.
\boxtheorem
\end{example}

A Datalog query $\mc{Q}(\bar{x})$ is a whole program $\Pi$ consisting of positive rules that accesses an underlying extensional database $E$ that is not part of the query. Program $\Pi$ contains a rule that defines
a top answer-collecting predicate $\nit{Ans}(\bar{x})$, by means of a rule of the form $\nit{Ans}(\bar{x}) \leftarrow P_1(\bar{s}_1), \ldots, P_m(\bar{s}_m)$. Now,  $\bar{a}$ is an answer to query $\Pi$ on $E$ when $\Pi \cup E \models \nit{Ans}(\bar{a})$. Here, entailment ($\models$) means that the RHS belongs to the minimal model of the LHS. So, the extension $\nit{Ans}(D)$ of $\nit{Ans}$ in the minimal model of the program contains the answers to the query.

A Datalog query is boolean if the top answer-predicate is propositional, say $\nit{ans}$, i.e. defined by a rule of the form $\nit{ans} \leftarrow P_1(\bar{s}_1), \ldots, P_m(\bar{s}_m)$. In this case, the query is true
if $\Pi \cup D \models \nit{ans}$, equivalently, if $\nit{ans}$ belongs to the minimal model of $\Pi \cup E$ \ \citep{ceri90,Abiteboul95}.

CQs can be expressed as Datalog queries, e.g. (\ref{eq:query}) becomes:

\vspace{-3mm}
{\small
\begin{eqnarray*}
\nit{Ans}_{\mc{Q}}(\sm{Name}, \sm{Topic})\!&\longleftarrow&\!\sm{Author}(\sm{Name,JName}),\nonumber\\
&& \sm{Journal}(\sm{JName,Topic,\#Paper}).
\end{eqnarray*}
}

\vspace{-3mm}
The definition of query-answer causality can be applied without any conceptual changes to Datalog queries.  In the case of Datalog, we sometimes use the notation $\nit{Causes}(E,\Pi(\bar{a}))$ or $\nit{Causes}(E,\Pi)$, depending on whether $\Pi$ has a $\nit{Ans}(\bar{x})$ or $\nit{ans}$ as answer predicate, resp.

In \citep{Meliou2010a}, causality for non-query answers is defined on basis of sets of {\em potentially missing tuples} that account for the missing answer. Computing actual causes and their responsibilities for
non-answers becomes a rather simple variation of causes for answers. In this work we focus on causality for query answers.

The complexity of the computational and decision problems that arise in query causality have been investigated in \citep{Meliou2010a,icdt15}. Here we present some problems and results that we use throughout this paper. The first is the causality problem, about deciding whether a tuple is an actual cause for a query answer.
\begin{definition}  \label{def:cp}    For a  boolean monotone query $\mc{Q}$,
the {\em causality decision problem} (CDP) is (deciding about membership of): \\ \vspace{1mm}
%\begin{equation}\label{eq:cpd}
$\mc{CDP}(\mc{Q}) := \{ (D,\tau)~|~ \tau \in D^n, \mbox{ and } \tau \in  \nit{Causes}(D,\mc{Q}) \}.$ \boxtheorem
\end{definition}

\vspace{-3mm}
This problem is tractable for UCQs \citep{icdt15}. The next is the responsibility problem, about deciding responsibility (above a given bound)  of a tuple for a query result.
\begin{definition}  \label{def:resp}
   For a  boolean monotone query $\mc{Q}$,
the {\em responsibility decision problem} (RDP) is (deciding about membership of): \\
\vspace{1mm}
 $\mathcal{RDP}(\mc{Q})=\{(D,\tau,v)~|~ \tau \in D^n, v \in \{0\} \ \cup $\\
 \hspace*{2cm} $\{\frac{1}{k}~|~k \in \mathbb{N}^+\},$  $D \models \mc{Q}$ \ and \ $\rho_{_{\!\mc{Q}\!}}(\tau) > v  \}$. \boxtheorem
\end{definition}

\vspace{-3mm}
This problem is \nit{NP}-complete for UCQs \citep{icdt15}, but tractable  for {\em linear} CQs \citep{Meliou2010a}. Roughly speaking, a CQ
is linear if its atoms can be ordered in a way that every variable appears in a continuous sequence of atoms that does not contain a self-join (i.e. a join involving the same predicate), e.g. $\exists xvyu(A(x) \wedge S_1(x, v) \wedge S_2(v, y) \wedge R(y, u) \wedge S_3(y, z))$ is linear, but not $\exists x y z(A(x) \wedge B(y) \wedge C(z) \wedge W(x, y, z))$, for which RDP is \nit{NP}-complete.  The class of CQs for which  RDP is tractable can be extended  to {\em weakly linear}.\footnote{Computing  sizes of minimum contingency sets is reduced to the max-flow/min-cut problem in a network.}

The functional, non-decision version of RDP, about computing the responsibility, i.e. an optimization problem, is complete for $\nit{FP}^{\nit{NP(log} (n))}$ for UCQs \citep{icdt15}.

%\combabak{The line in red is allright}

Finally, we have the problem of deciding weather a tuple is a most responsible cause:
\begin{definition}  \label{def:mracp}   For a  boolean monotone query $\mc{Q}$, the {\em most responsible cause decision problem} (MRDP)
 is:\\
 \vspace{1mm}
   $\mc{MRCD}(\mc{Q})$ $=\{(D,\tau)~|~ \tau \in D^n \ \mbox{ and }$\\
   \hspace*{3cm} $ 0 < \rho_\mc{Q}(\tau) \mbox{ is a maximum for } D\}$.
\boxtheorem
\end{definition}

\vspace{-3mm}
For UCQs this problem is complete for  $\nit{P}^{\nit{NP(log} (n))}$ \citep{icdt15}.

\vspace{-2mm}
\subsection{VIEW-CONDITIONED CAUSALITY}

\vspace{-2mm}
A form of {\em conditional  causality} was informally introduced in \citep{Meliou2010b},   to characterize  causes for a  query answer that are  conditioned by the other answers to the query.
The notion was made precise in \citep{Meliou2011}, in a more general, non-relational setting that in particular includes
the case of several queries. In them the notion of {\em view-conditioned causality was used}, and we adapt it in the following to the case of a single query, possibly with several answers.

\ignore{\combabak{ In \cite{Meliou2010b}[page 6] they dedicate a short subsection to the subject and motivate it. However, they don't have a full formal definition. On the other hand, in \cite{Meliou2011}  a formal definition of view conditioned causality has been provided for data transformations in general i.e., they go beyond relational databases.}  }

 Consider an instance $D=D^n \cup D^x$, and a monotone query $\mc{Q}$ with $\mc{Q}(D)=\{\bar{a}_1,\ldots \bar{a}_n\}$. Fix an answer, say $\bar{a}_k \in \mc{Q}(D)$, while the other answers  will be used as a condition on
 $\bar{a}_k$'s causality. Intuitively, $\bar{a}_k$ is somehow unexpected, and we look for causes, by considering the other answers as ``correct". The latter assumption has, in technical terms, the effect of
 reducing the spectrum of contingency sets, by keeping $\mc{Q}(D)$'s extension fixed, as a view, modulo the answer  $\bar{a}_k$ at hand.

 \begin{definition}  \label{def:vccause} \begin{enumerate}[(a)] \item  A tuple $\tau \in D^n$ is called a {\em view-conditioned counterfactual cause} (VCC-cause) for answer $\bar{a}_k$ to $\mc{Q}$ if  $D \smallsetminus \{\tau\} \not \models \mc{Q}(\bar{a}_k)$ and  $D \smallsetminus\{\tau\} \models \mc{Q}(\bar{a}_i)$, for \ $ i \in \{1, \ldots,n\} \smallsetminus \{k\}$. \item A tuple $\tau \in D^n$  is an {\em view-conditioned actual cause} (VC-cause) for  $\bar{a}_k$ if there  exists a contingency set, $\Gamma \subseteq D^n$,
 such $\tau$ is a  VCC-cause for $\bar{a}_k$ in $D \smallsetminus \Gamma$.
 \item $\nit{vc\mbox{-}\!Causes}(D,\mc{Q}(\bar{a}_k))$ denotes the set of all VC causes for $\bar{a}_k$.
 \boxtheorem
 \end{enumerate}
\end{definition}

\vspace{-4mm}
 Intuitively, a tuple $\tau$ is a VC-cause for $\bar{a}_k$ if there is a contingent state of the database that entails all the answers to $\mc{Q}$ and $\tau$ is a counterfactual cause for $\bar{a}_k$, but not for the rest of the answers. Obviously, VC-causes for $\bar{a}_k$ are also actual causes, but not necessarily the other way around:  $ \nit{vc\mbox{-}\!Causes}(D,Q(a_k)) \subseteq \nit{Causes}(D,Q(a_k))$.

\begin{example}\label{ex:cfex2} (ex. \ref{ex:cfex1} cont.) Consider the same instance $D$, query $\mc{Q}$, and the answer $\langle \sm{John, XML}\rangle$, which does not have any VC-cause. To see this, take for example, the tuple \sm{ Author(John, TODS)} that is an actual cause for $\langle \sm{John, XML}\rangle$, with two contingency sets, $\Gamma_1$ and $\Gamma_2$. It is easy to verify that none of these contingency sets satisfies the condition in Definition \ref{def:vccause}, e.g. the original answer $\langle \sm{John, CUBE}\rangle$ is not such anymore from $D \smallsetminus \Gamma_1$. The same argument can be applied to all actual causes for $\langle \sm{John, XML}\rangle$. \boxtheorem
\end{example}

\vspace{-3mm}
This example shows that it makes sense to study the complexity of deciding whether a query answer has a VC-actual cause or not.

\begin{definition}  \label{def:mracpV}
  For a monotone query $\mc{Q}$, the {\em view-conditioned cause  problem} is (deciding about membership of):\\
  \vspace{1mm}
   $\mc{VCP}(\mc{Q})$ $=\{(D,\bar{a})~|~  \bar{a} \in \mc{Q}(D) \ \mbox{ and}$\\  \hspace*{3.5cm}$\nit{vc\mbox{-}\!Causes}(D, \mc{Q}(\bar{a})) \not = \emptyset \ \}.$
\boxtheorem
\end{definition}

\vspace{-3mm}

\vspace{-5mm}
\section{CAUSALITY AND ABDUCTION} \label{sec:abdandcause}

\vspace{-2mm}
In general  logical terms, an abductive explanation of an observation is a formula that, together with the background logical theory, entails the observation. So, one could see an abductive explanation as
a cause for the observation. However, it has been argued that causes and abductive explanations are not necessarily the same \citep{Psillos96,DeneckerK02}. \ignore{A classic example from \citep{Psillos96} is as follows:  the disease
{\em paresis} is caused by a latent untreated form of {\em syphilis}. The probability that latent untreated syphilis leads
to paresis is only 25\%. Therefore, syphilis is the cause of paresis but does not entail it, while
paresis entails syphilis but does not cause it.}

Under the abductive approach to diagnosis \citep{console91,EiterGL95,Poole92,Poole94}, it is common that the system specification rather explicitly describes causality information,
specially in action theories where the
effects of actions are directly represented by Horn formulas. By restricting the explanation formulas to the predicates describing primitive
causes (action executions), an explanation formula which entails an observation gives also a cause for the observation \citep{DeneckerK02}. In this
case, and is some sense, causality information is imposed by the system specifier \ \citep{Poole92}.

In database causality we do not have, at least not initially, a system description,\footnote{Having integrity constraints would go in that direction, but we are not considering
their presence in this work. However, see \cite[sec. 5]{icdt15} for a consistency-based diagnosis connection.} but just a set of tuples. It is when we pose a query that we create something like a description, and the causal relationships between tuples are captured
by the combination of  atoms in the query. If the query is a Datalog query (in particular, a CQ), then we have a Horn specification too.

In this section we will establish
 connections between abductive diagnosis and database causality.\footnote{In \citep{icdt15} we established such a connection
 between another form of model-based diagnosis \citep{struss}, namely consistency-based diagnosis \citep{Reiter87}. For relationships
and comparisons between consistency-based and abductive diagnosis see \citep{console91}.} For that, we have to be more precise about the kind of abduction problems we will consider.

\vspace{-2mm}
\subsection{BACKGROUND ON DATALOG ABDUCTIVE DIAGNOSIS}\label{sec:backAbd}

\vspace{-2mm}

A  {\em Datalog abduction problem} \citep{EiterGL97} is of the form $\mathcal{AP}= \langle \Pi, E, \nit{Hyp},  \nit{Obs}\rangle$, where: \ (a) $\Pi$ is a set of Datalog rules,
 (b) $E$ is a  set of ground atoms (the extensional database), whose predicates do not appear in heads of rules in $\Pi$, (c) $\nit{Hyp}$, the hypothesis, is a finite set of
ground atoms, the abducible atoms in this case,\footnote{It is common to accept as hypothesis all the possible ground instantiations of {\em abducible predicates}. We assume abducible
predicates do not appear in rule heads.} and (d) $\nit{Obs}$, the observation, is a finite conjunction of
ground atoms. As it is common, we will start with the assumption that $\Pi \cup E \cup \nit{Hyp} \models \nit{Obs}$.

The {\em abduction problem} is about computing a minimal $\Delta \subseteq \nit{Hyp}$ (under certain minimality criterion), such that $\Pi \cup E \cup \Delta \models \nit{Obs}$.
More specifically:
\begin{definition} Consider  a {\em Datalog abduction problem} $\mathcal{AP}= \langle \Pi, E, \nit{Hyp},  \nit{Obs}\rangle$
\begin{enumerate}[(a)] \vspace{-.3cm}\item
 An  {\em  abductive diagnosis} (or simply, {\em a solution}) for $\mathcal{AP}$ is a subset-minimal $\Delta \subseteq \nit{Hyp}$, such that
$ \Pi \cup E \cup \Delta \models \nit{Obs}$. This requires that no proper subset of $\Delta$ has this property.
 $\nit{Sol}(\mathcal{AP})$ denotes the set of abductive diagnoses for problem $\mc{AP}$.
\vspace{-.1cm} \item
A hypothesis $h \in \nit{Hyp}$  is {\em relevant} for $\mathcal{AP}$ if  $h$ contained in at least one diagnosis of $\mathcal{AP}$. $\nit{Rel}(\mc{AP})$ collects  all relevant hypothesis for $\mc{AP}$.
\boxtheorem
\end{enumerate}
\end{definition}

\vspace{-3mm}We are interested in deciding, for a fixed Datalog program, if an hypothesis is relevant or not, with all the data as input.

 More precisely, we consider the following decision problem.
\begin{definition}  \label{def:relp}  Given a Datalog program $\Pi$, the {\em relevance decision problem} (RLDP) {\em for} $\Pi$ is (deciding about the membership of):\\
\vspace{1mm}$\mc{RLDP}(\Pi)=\{ (\nit{E,Hyp,Obs}, h)~|~ h \in \nit{Rel}(\mc{AP}), \mbox{ with } $\\
\hspace*{2cm} $\mc{AP}= \langle \Pi, E, \nit{Hyp},  \nit{Obs}\rangle \mbox{ and } h \in \nit{Hyp} \}$.
\boxtheorem
\end{definition}

\vspace{-3mm}As it is common, we will assume that $|\nit{Obs}|$, i.e. the number of atoms in the conjunction, is bounded above by a numerical parameter $p$. It is common that $p=1$ (a single atomic
observation). %\red{ We will also assume that $|\nit{Hyp}|$ is polynomially bounded by $|E|$.}

Definition \ref{def:relp} suggests that we are interested in the {\em data complexity} of the relevance problem for Datalog abduction. That is, the Datalog program is fixed and
hypotheses and input structure may change and maybe regarded as data. In contrast, under {\em combined complexity} the program is also part of the input, and the complexity is measured also
in terms of the program size.

The following result is obtained by showing that the  $\nit{NP}$-complete combined complexity of the relevance problem for Propositional Datalog Abduction (PDA) (established in \citep{Friedrich90}), coincides with the data complexity of the relevance problem for (non-propositional) Datalog Abduction. For this, techniques developed in \citep{EiterGL97} can be used.

\begin{proposition}\label{pro:relp} \em For every Datalog program $\Pi$,
$\mc{RLDP}(\Pi) \in \nit{NP}$, and there are programs $\Pi'$ for which $\mc{RLDP}(\Pi')$
 is {\em NP}-hard.
 \boxtheorem
\end{proposition}

\vspace{-3mm}

It is clear from this result that the combined complexity of deciding relevance for Datalog abduction is also intractable.  However, a tractable case of combined complexity is identified  in \citep{Gottlob10b}, on the basis of the notions of {\em tree-decomposition and bounded tree-width}, which we now briefly present.

Let $\mc{H} = \langle V,H \rangle$ be a hypergraph. $V$ is the set of vertices, and $H$ the set of hyperedges, i.e. of subsets of $V$. A tree-decomposition $\mc{T}$ of $\mc{H}$ is a pair $(\mc{T}, \lambda)$, where $\mc{T} = \langle N, E\rangle$ is a tree and $\lambda$ is a labeling
function that assigns to each node $n \in N$, a subset  $\lambda(n)$ of  $V$ ($\lambda(n)$ is aka. bag), i.e. $\lambda(n) \subseteq V$, such that, for every node $n \in N$, the following hold: \ (a) For every $v \in V$ , there exists $n \in N$ with
$v \in \lambda(n)$. \
(b) For every $h \in H$, there exists a node $n \in N$ with $h \subseteq \lambda(n)$. \ (c) For every $v \in V$, the set of nodes $\{n ~|~ v \in \lambda(n)\}$ induces a connected subtree of $\mc{T}$.

The {\em width} of a tree decomposition $(\mc{T}, \lambda)$ of $\mc{H} = \langle V,H \rangle$, with $\mc{T} = \langle N, E\rangle$, is defined as $\nit{max}\{|\lambda(n)| - 1 \ : \  n \in N\}$.  The tree-width $t_w(\mc{H})$ of $\mc{H}$ is
the minimum width over all its tree decompositions.

Intuitively, the tree-width of a hypergraph $\mc{H}$ is a
measure of the ``tree-likeness" of $\mc{H}$. A set of vertices that form a cycle in $\mc{H}$ are put into a same bag, which becomes (the bag of a) node in the corresponding tree-decomposition.
If the tree-width of the
hypergraph under consideration is bounded by a fixed constant,
then many otherwise intractable problems become tractable \citep{Gottlob10}.

It is possible to associate an hypergraph to any finite structure $D$ (think of a relational database): If its universe (the active domain in the case of a relational database) is $V$, define the hypergraph  $\mc{H}(D) = (V,H)$, with
$H = \{~\{a_1, \ldots , a_n\}~ | ~ D$ contains a ground atom $P(a_1 \ldots a_n)$ for some predicate symbol $P \}$.

\begin{figure}
  \centering
  \epsfig{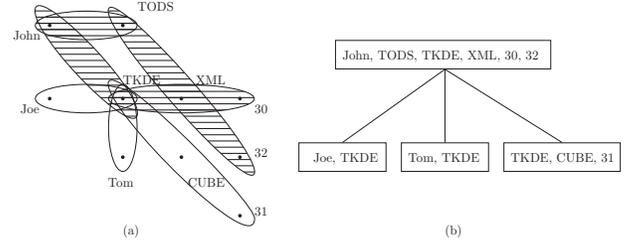}\vspace{-5mm}
  \caption{ \ (a) \ $\mc{H}(D)$. \  \ (b) \ A  tree decomposition of $\mc{H}(D)$. }\label{fig:fig1}\vspace{-8mm}
\end{figure}

\begin{example}\label{ex:cfex3} Consider instance $D$ in Example \ref{ex:cfex1}. The hypergraph $\mc{H}(D)$ associated to $D$ is shown in Figure \ref{fig:fig1}(a). Its vertices are the elements of
$\nit{adom}(D) = \{\nit{John, Jone, Tom, TODS, TKDE, XML, Cube, 30, 31,}$ $\nit{ 32 } \}$, the active domain of $D$. For example, since
$\nit{Journal(TKDE, XML, 30)} \in D$, $\{\nit{TKDE, XML, 30}\}$ is one of the hyperedges.

The dashed ovals show four sets of vertices, i.e. hyperedges,  that together form a cycle. Their elements are put into the same bag of the tree-decomposition. Figure \ref{fig:fig1}(b) shows a possible tree-decomposition of $\mc{H}(D)$.
In it, the maximum $|\lambda(n)| -1$ is $6 -1$, corresponding to the top box bag of the tree. So, $t_w(\mc{H}(D)) \leq 5$. \boxtheorem \end{example}

\vspace{-5mm}
The following is a {\em fixed-parameter tractability} result for the relevance decision problem for Datalog abduction problems
with a program $\Pi$ that is {\em guarded}, which means that in every rule body there
is an atom that contains (guards) all the variables appearing in that body.

\begin{theorem} \label{the:trac} \em  \citep{Gottlob10b} \
Let $k$ be an integer. For Datalog abduction problems $\mathcal{AP}=\langle \Pi, E,
\nit{Hyp},\nit{Obs}\rangle$ where $\Pi$ is guarded, and
$t_w(\mc{H}(E)) \leq k$, relevance can be decided in polynomial time in $|\mc{AP}|$.\footnote{This is Theorem 7.9 in \citep{Gottlob10b}.} More precisely, the decision problem: \
$\mc{RLDP}=\{ (\langle \Pi,\nit{E,Hyp,Obs}\rangle, h)~|~ h \in \nit{Rel}(\langle \Pi,\nit{E,Hyp,Obs}\rangle),
 h \in \nit{Hyp}, \Pi \mbox{ is guarded, and } t_w(\mc{H}(E)) \leq k \}$ is tractable.
\boxtheorem
\end{theorem}

\vspace{-5mm}
This is a case of tractable combined complexity with a  fixed parameter that is the
tree-width of the extensional database.

\subsection{QUERY CAUSALITY FROM ABDUCTIVE DIAGNOSIS} \label{sec:abdcomx}

\vspace{-2mm}
In this section we first show that, for the class of Datalog theories (system specifications),  abductive inference corresponds to actual causation for monotone queries.
That is,  abductive diagnoses for an observation essentially contain actual causes for the observation.

Assume that $\Pi$ is a boolean, possibly recursive Datalog query. Consider the relational instance  $D=D^x \cup D^n$.
Also assume that $\Pi \cup D \models \nit{ans}$. So, the decision problem in Definition \ref{def:cp} takes the form
$\mc{CDP}(\Pi) := \{ (D,\tau)~|~ \tau \in D^n, \mbox{ and } \tau \in  \nit{Causes}(D,\Pi) \}$.

We now show that  actual causes for $\nit{ans}$ can be obtained from abductive diagnoses of the associated {\em causal Datalog abduction problem} (CDAP): \  $\mathcal{AP}^c:=\langle \Pi,  D^x, D^n,\nit{ans}\rangle$, where
 $D^x$ is the extensional database for $\Pi$ (and then $\Pi \cup D^x$ becomes  the {\em background theory}),  $D^n$ becomes the set of {\em hypothesis}, and atom $\nit{ans}$ is the observation.
\begin{proposition} \label{pro:abdf&cfcaus}  \em
$t \in D^n$ is an actual cause for $\nit{ans}$  iff $t \in \nit{Rel}( \mathcal{AP}^c)$.
 \boxtheorem
\end{proposition}

\vspace{-5mm}
\begin{example} \label{ex:abdex3}
Consider the instance $D$ with relations $R$ and $S$ as below, and the query $\Pi\!: \ \nit{ans} \leftarrow R(x, y), S(y)$,
which is true in $D$.
Assume all tuples are endogenous.

\begin{center} \begin{tabular}{l|c|c|} \hline
$R$  & X & Y \\\hline
 & $a_1$ & $a_4$\\
& $a_2$ & $a_1$\\
& $a_3$ & $a_3$\\
 \hhline{~--}
\end{tabular} \hspace*{1cm}\begin{tabular}{l|c|c|}\hline
$S$  & X  \\\hline
 & $a_1$ \\
& $a_2$ \\
& $a_3$ \\ \hhline{~-}
\end{tabular}
\end{center}

$\mathcal{AP}^c= \langle \Pi,\emptyset, D, \nit{ans}\rangle$
has two (subset-minimal) abductive diagnoses:  $\Delta_1=\{S(a_1),  R(a_2, a_1) \}$ and $\Delta_2=\{S(a_3),  R(a_3, a_3)\}$. Then,
$\nit{Rel}(\mc{AP}^c)=\{ S(a_3),$ $  R(a_3, a_3),$ $S(a_1),  R(a_2, a_1)\}$.
It is easy to see that the relevant hypothesis are actual causes for $\nit{ans}$.
\boxtheorem
\end{example}

\vspace{-5mm}
We are  interested in obtaining {\em responsibilities} of actual causes for $\nit{ans}$.
\begin{definition} \label{def:ness}
Given a CDAP, $\mathcal{AP}^c=\langle \Pi,D^x,D^n,\nit{ans}\rangle$, with  $\nit{Sol}(\mathcal{AP}^c)\neq \emptyset$, \ $N \subseteq D^n$ is a {\em necessary-hypothesis set} if $N$ is  subset-minimal such that
$\nit{Sol}(\mathcal{AP}^c_N)=\emptyset$, with $\mathcal{AP}^c_N :=\langle \Pi,D^x,D^n  \smallsetminus N,\nit{ans}\rangle$.
\boxtheorem
\end{definition}
\vspace{-5mm}
\begin{proposition} \label{pro:abdf&res}  \em
The responsibility of a tuple $t$ for $\nit{ans}$ is $\frac{1}{|N|}$, where $N$ is a necessary-hypothesis set with minimum cardinality for $\mathcal{AP}^c$ and $t \in N$.
 \boxtheorem
\end{proposition}
\vspace{-5mm}
In order to represent Datalog abduction in terms of query-answer causality, we show that abductive diagnoses from Datalog programs  are formed essentially by actual causes for the observation.

More precisely, consider a Datalog abduction problem $\mathcal{AP}= \langle \Pi, E, \nit{Hyp},  \nit{Obs}\rangle$, where $E$ is the
underlying extensional database, and
$\nit{Obs}$ is a conjunction of ground atoms.

Now we construct a
query-causality setting: $D := D^x \cup D^n$, $D^x:= E$, and $D^n:= \nit{Hyp}$. Consider the program
 $\Pi' := \Pi \cup \{\nit{ans} \leftarrow \nit{Obs}\}$ (with \nit{ans} a fresh propositional atom). So, $\Pi'$  is seen as a monotone query on $D$.

\begin{proposition} \label{pro:ac&rel}  \em
A hypothesis $h$ is relevant for $\mathcal{AP}$, i.e. $h \in \nit{Rel}( \mathcal{AP})$, iff $h$ is an actual cause for \nit{ans} wrt. $\Pi',D$.
 \boxtheorem
\end{proposition}
\vspace{-5mm}

Now we will use the results obtained so far in this section to obtain new complexity results for Datalog  query causality.
 Actually, the following result is obtained from  Propositions \ref{pro:relp} and  \ref{pro:abdf&cfcaus}:

\begin{proposition}\label{pro:cp} \em
For boolean Datalog queries $\Pi$, $\mc{CDP}(\Pi)$ is {\em NP}-complete (in data).
 \boxtheorem
\end{proposition}
\vspace{-5mm}
This result should be contrasted with the tractability of same problem for UCQs \citep{icdt15}.

We now introduce a fixed-parameter tractable case of this problem. For this we take advantage of the tractable case of Datalog abduction presented in Section \ref{sec:backAbd}.
The following is a consequence of Theorem \ref{the:trac} and  Proposition \ref{pro:abdf&cfcaus}.

\begin{proposition}\label{cor:traccaus} \em
For guarded Datalog queries $\Pi$ and a extensional instances $D = D^x \cup D^n$, with $D^x$ of bounded tree-width, $\nit{CDP}$ is fixed-parameter tractable in combined complexity, with
the parameter being the tree-width bound.
\boxtheorem
\end{proposition}

\vspace{-4mm}
\section{VIEW-UPDATES AND QUERY CAUSALITY} \label{sec:delp&cause}

\vspace{-3mm}
There is a close relationship between query causality  and the view-update problem in the form of delete-propagation, which was first suggested in  \citep{Kimelfeld12a,Kimelfeld12b} (see
also \citep{BunemanKT02}). We start by formalizing some specific computational problems related to the general delete-propagation problem.

\vspace{-1mm}
\subsection{DELETE-PROPAGATION PROBLEMS}\label{sec:del-pro}

\vspace{-2mm}
Given a monotone query $\mc{Q}$, we can think of it as defining a view with virtual contents $\mc{Q}(D)$. If $\bar{a} \in \mc{Q}(D)$, which may not be intended,
we  may try to delete some tuples from $D$, so that  $\bar{a}$  disappears from $\mc{Q}(D)$. This is a common case of the problem of database updates through views
\citep{Abiteboul95}.
 In this work we consider some variations of this problem, in both their functional and the decision versions.

\begin{definition}  \label{def:SubSDP}  For an instance $D$, and a monotone query $\mc{Q}$: \vspace{-3mm}
\begin{enumerate}[(a)]
\item For $\bar{a} \in \mc{Q}(D)$, the {\em minimal source-side-effect problem} is about computing a subset-minimal $\Lambda \subseteq D$, such that $\bar{a} \ \notin \mc{Q}(D \smallsetminus \Lambda)$.

\vspace{-2mm}
\item The {\em minimal source-side-effect decision problem}  is (deciding about the membership \ of):

 $\mc{MSSEP}^{s\!}(\mc{Q})= \{(D,D',\bar{a}) ~|~ \bar{a}\in \mc{Q}(D), \ D' \subseteq D,$$\\
 \hspace*{1.7cm}$ $ \  \bar{a} \not \in \mc{Q}(D'), \mbox{ and }  D' \mbox{ is subset-maximal} \}$.\\
 {\small (The superscript $s$ stands for subset-minimal.)}

\vspace{-2mm}
\item For $\bar{a} \in \mc{Q}(D)$, the {\em minimum} {\em source}  {\em side-effect problem} is about computing a minimum-cardinality $\Lambda \subseteq D$, such that $\bar{a} \notin \mc{Q}(D \smallsetminus \Lambda)$.

\vspace{-2mm}
\item The {\em minimum} {\em source}  {\em side-effect} {\em decision problem} is (deciding about the membership of):

$\mc{MSSEP}^{c\!}(\mc{Q})= \{(D,D',\bar{a}) ~|~ \bar{a}\in \mc{Q}(D), D' \subseteq D,$\\ \hspace*{0.9cm} $\bar{a} \notin \mc{Q}(D'), \mbox{ and } D' \mbox{ has maximum cardinality} \}$.\\
{\small (Here $c$ stands for minimum cardinality.)}
\boxtheorem
\end{enumerate}
\end{definition}

\vspace{-5mm}
\begin{definition}  \label{def:FreeVDP} \citep{BunemanKT02} \ For an instance $D$, and a  monotone query $\mc{Q}$:   \vspace{-3mm}
\begin{enumerate}[(a)]
\item For $\bar{a} \in \mc{Q}(D)$, the {\em view side-effect-free problem} is about computing a  \ignore{minimum-cardinality} $\Lambda \subseteq D$, such that $\mc{Q}(D) \smallsetminus \{\bar{a}\} = \mc{Q}(D
\smallsetminus \Lambda)$.
\vspace{-3mm}
\item The {\em
view side-effect-free decision problem} is (deciding
about the membership of):

$\mc{VSEFP}(\mc{Q})= \{(D,\bar{a}) ~|~  \bar{a} \in
\mc{Q}(D), \ \mbox{ and exists }$ \\ \hspace*{1.7cm}$D' \subseteq D \mbox{ with } \mc{Q}(D) \smallsetminus
\{\bar{a}\} = \mc{Q}(D')\} $. \boxtheorem
\end{enumerate}
\end{definition}

\vspace{-7mm}
\subsection{VIEW DELETIONS VS. CAUSES}

\vspace{-2mm}
In this section we first establish mutual reductions between the different variants of the delete propagation problem and both query and view-conditioned
causality. On this basis, we obtain next some complexity results for view-conditioned causality and the minimum source-side-effect
problem.

{\em In this section all tuples in the instances involved are assumed to be endogenous.} \
Consider a relational database $D$, a view $ \mc{V}$ defined by a monotone query $\mc{Q}$. So, the virtual view extension, $\mc{V}(D)$, is $\mc{Q}(D)$.

For a tuple $\bar{a} \in \mc{V}(D)$, the delete-propagation problem, in its most general form, is the task of deleting a set of tuples from $D$, and so obtaining a subinstance $D'$ of $D$, such that
$\bar{a} \notin \mc{V}(D')$. It is natural to expect that the deletion of $\bar{a}$ from the view
can be achieved through deletions from $D$ of the causes for $\bar{a}$ to be in the view extension.
However, to obtain solutions to the different variants of this problem introduced in Section \ref{sec:del-pro}, different sets of actual causes must be considered.

First, we show that an actual cause for $\bar{a}$ to be in $\mc{V}(D)$ forms, with any of its contingency sets,  a solution to the minimal source-side-effect problem (cf. Definition \ref{def:SubSDP}).

\begin{proposition} \label{pro:causeSubSDP} \em
 Consider an instance $D$, a view $ \mc{V}$ defined by a monotone query $\mc{Q}$, and  $\bar{a} \in \mc{V}(D)$: \ $D' \subseteq D$ is a solution to the minimal source-side-effect problem, i.e. $(D,D',\bar{a}) \in \mc{MSSEP}^{s\!}(\mc{Q})$, \ iff \ there is a $t \in D \smallsetminus D'$, such that  $ t\in\nit{Causes}(D,\mc{Q}(\bar{a}))$ and $ D \smallsetminus (D' \cup  \{t\}) \in \nit{Cont}(D,\mc{Q}(\bar{a}), t)$.
\boxtheorem
\end{proposition}
\vspace{-5mm}
Now  we show that, in order to minimize the side-effect on the source (cf. Definition \ref{def:SubSDP}(c)),
it is good enough to pick a most responsible cause for $\bar{a}$ with any of its minimum-cardinality contingency sets.
\begin{proposition} \label{pro:causeminSDP} \em
 Consider an instance $D$, a view $ \mc{V}$ defined by a monotone query $\mc{Q}$, and  $\bar{a} \in \mc{V}(D)$: \ $D' \subseteq D$ is a solution to the minimum source-side-effect problem, i.e. $(D,D',\bar{a}) \in
\mc{MSSEP}^{c\!}(\mc{Q})$, \ iff \ there is a $t \in D \smallsetminus D'$, such
that  $t \in  \mc{MRC}(D,\mc{Q}(\bar{a}))$, $ \Lambda:= D \smallsetminus (D'
\cup \{t\}) \in \nit{Cont}(D,\mc{Q}(\bar{a}), t)$, and there is no $\Lambda'
\in \nit{Cont}(D,\mc{Q}(\bar{a}), t)$ with $|\Lambda'|< |\Lambda|$.
\boxtheorem
\end{proposition}
\vspace{-5mm}
Next, we show that in order to check if there exists a solution to the view side-effect-free problem for  $\bar{a} \in \mc{V}(D)$ (cf. Definition \ref{def:FreeVDP}), it is good enough to check if $\bar{a}$ has a view-conditioned cause.\footnote{Since this proposition does not involve contingency sets, the existential problem in Definition \ref{def:FreeVDP}(b) is the right one to consider.}
\begin{proposition} \label{pro:VC&view} \em
 Consider an instance $D$, a view $ \mc{V}$ defined by a monotone query $\mc{Q}$, and  $\bar{a} \in \mc{V}(D)$: \ There is a solution to the view side-effect-free problem for $\bar{a}$, i.e. $(D,\bar{a}) \in \mc{VSEFP}(\mc{Q})$, \ iff \
 $\nit{vc\mbox{-}\!Causes}(D,\mc{Q}(\bar{a})) \not = \emptyset$.
\boxtheorem
\end{proposition}
\vspace{-3mm}
\begin{example}\label{ex:Vpex1} (ex. \ref{ex:cfex1} cont.) Consider the same instance $D$,  query $\mc{Q}$, and answer $\langle \sm{ John, XML}\rangle$.

Consider the following sets of tuples:

$S_1$=\{ \sm{ Author(John, TKDE), Journal(TODS, XML, 32)}\},

$S_2$=\{ \sm{ Author(John, TODS), Journal(TKDE, XML, 30)}\},

$S_3$=\{ \sm{ Journal(TODS, XML, 30), Journal(TKDE, XML, 30)}\},

$S_4$=\{ \sm{Author(John, TODS), Author(John, TKDE)}\}.

Each of the subinstances $D \smallsetminus S_i$, $i=1,\ldots,4$,  is a solution to both the minimum and minimal source-side-effect problems. These solutions essentially contain the actual causes for answer $\langle\sm { John, XML}\rangle$, as computed in Example \ref{ex:cfex1}. Moreover, there is no solution to the view side-effect-free problem associated to this answer, which coincides with the result obtained in  Example  \ref{ex:cfex2}, and confirms Proposition \ref{pro:VC&view}.
\boxtheorem
\end{example}

\vspace{-5mm}
Now we show, the other way around, that  actual causes, most responsible causes, and VC causes can be obtained from solutions to different variants of the delete-propagation problem.

First, we show that actual causes for a query answer can be obtained from the solutions to the corresponding minimal source-side-effect problem.
\begin{proposition} \label{pro:causefromview} \em
 Consider an instance $D$, a view $ \mc{V}$ defined by a monotone query $\mc{Q}$, and  $\bar{a} \in \mc{V}(D)$: \
Tuple $t$ is an actual cause for $\bar{a}$ \ iff \ there is a $D' \subseteq D$ with $t\in  (D\smallsetminus D') \subseteq D^n$
and $(D,D',\bar{a}) \in \mc{MSSEP}^{s\!}(\mc{Q})$.
\boxtheorem
\end{proposition}
\vspace{-5mm}
Similarly, most-responsible causes for a query answer can be obtained from solutions to the corresponding minimum source-side-effect problem.
\begin{proposition} \label{pro:mostrescausefromview} \em
 Consider an instance $D$, a view $ \mc{V}$ defined by a monotone query $\mc{Q}$, and  $\bar{a} \in \mc{V}(D)$: \
 Tuple $t$ is a most responsible actual cause for $\bar{a}$ \ iff \ there is a $D' \subseteq D$ with $t \in (D\smallsetminus D') \subseteq D^n$  and  $(D,D',\bar{a}) \in \mc{MSSEP}^{c\!}(\mc{Q})$.
\boxtheorem
\end{proposition}
\vspace{-5mm}
Finally, VC-causes for an answer can be obtained from solutions to the view side-effect-free problem.
\begin{proposition} \label{pro:VCcausefromview} \em
 Consider an instance $D$, a view $ \mc{V}$ defined by a monotone query $\mc{Q}$, and  $\bar{a} \in \mc{V}(D)$: \
Tuple $t$ is a
VC-cause for $\bar{a}$ \ iff \ there is a  $D' \subseteq D$ with $t \in
(D\smallsetminus D') \subseteq D^n$  and $D'$ is a solution to the view side-effect-free problem associated to $\bar{a}$.
\boxtheorem
\end{proposition}

\vspace{-5mm}
The partition of a database into endogenous and exogenous tuples
used in causality may also be of interest in the context of delete propagation. It makes sense to consider {\em endogenous delete-propagation} that are obtained through deletions
on endogenous tuples only. Actually, given an instance $D=D^n \cup D^x$, a view $ \mc{V}$ defined by a monotone query $\mc{Q}$, and  $\bar{a} \in \mc{V}(D)$, endogenous delete-propagations for  $\bar{a}$ (in all of its flavors)  can be obtained from actual causes for $\bar{a}$ from the partitioned instance.

%\vspace{-3mm}
\begin{example} \label{exa:dppar} \ (ex. \ref{ex:Vpex1} cont.) \ Consider again that tuple $\langle\sm{ John, XML}\rangle$ must be deleted from the query result; and assume now the data in \sm{Journal} is reliable. Therefore,  only deletions from \sm{Author} make sense.  This can be captured by considering \sm{Journal}-tuples as exogenous and \sm{Author}-tuples as endogenous. With this partitioning, only $ \sm{\small Author(John, TODS)}$ and $\sm{ Author(John, TKDE)}$ are  actual causes for $\langle\sm{ John, XML}\rangle$, and each of them  forms a singleton and unique contingency set of the other as a cause (See Example{ex:cfex1}). Therefore,
$D \smallsetminus\{ \sm{\small Author(John, TODS), Author(John, TKDE)}\}$ is a solution to the associated minimal- and minimum endogenous delete-propagation of  $\langle\sm{ John, XML}\rangle$.\boxtheorem
\end{example}

\vspace{-5mm}
We now investigate the complexity of the view-conditioned causality problem  (cf. Definition \ref{def:mracpV}). For this, we take advantage of the connection between VC-causality and the view side-effect-free  problem.
Actually, the following result is obtained from the \nit{NP}-completeness of view side-effect-free  problem \citep{BunemanKT02} and Proposition \ref{pro:VC&view}.
\begin{proposition} \label{pro:causefromviewII} \em
For CQs,  the view-conditioned causality decision problem, $\mc{VCP}$, is  \nit{NP}-complete.
\boxtheorem
\end{proposition}

%\comlb{I guess this also holds for UCQs?}

%\combabak{Your conjecture is true}

\vspace{-5mm}
Actually, this result also holds for UCQs.
The next result is obtained from  the $\nit{FP}^\nit{NP(log(n))}$-completeness of computing the responsibility of the most
responsible causes (obtained in \citep{icdt15}) and Proposition \ref{pro:causeminSDP}.
\begin{proposition} \label{pro:causeview} \em
Computing the size of a solution to the minimum source-side-effect problem
 is $\nit{FP}^\nit{NP(log(n))}$-hard. \boxtheorem
\end{proposition}

\vspace{-5mm}
As mentioned in Section \ref{sec:causeintro}, responsibility computation (more precisely the RDP problem in Definition \ref{def:resp}) is tractable for weakly linear queries. We can take advantage of this result and obtain,
via Proposition \ref{pro:causeminSDP}, a new tractability result for the minimum source-side-effect problem, which has been shown to be \nit{NP}-hard  for general CQs in \citep{BunemanKT02}.

\begin{proposition} \label{pro:dichsse} \em
For weakly linear queries, the minimum source-side-effect decision problem is tractable.\boxtheorem
\end{proposition}

\vspace{-5mm}The class of weakly linear queries generalizes that of linear queries (cf. Section \ref{sec:causeintro}). So, Proposition \ref{pro:dichsse} also holds for linear queries.

In \citep{BunemanKT02} it has been shown that the minimum source-side-effect decision problem is tractable for the class of
project-join queries with {\em  chain joins}. Now, a join on $k$ atoms with different predicates, say
$R_1,...,R_k$, is a chain join if there are no  attributes (variables) shared by
any two atoms $R_i$ and $R_j$ with $j > i + 1$. That is, only consecutive relations may share attributes. For example,   $\exists xvyu(A(x) \wedge S_1(x, v) \wedge S_2(v, y) \wedge R(y, u) \wedge S_3(y, z))$ is a project-join query with chain joins.

We observe that project-join queries with chain joins correspond
linear queries. Actually, the tractability results for these classes of queries are both obtained via a reduction to maximum flow problem \citep{Meliou2010a,BunemanKT02}.  As a consequence, the result in Proposition \ref{pro:dichsse} extends that in \citep{BunemanKT02}, from linear queries to weakly-linear queries. For example,  $\exists xyz( R(x, y) \wedge S(y, z) \wedge T(z, x) \wedge V(x))$ is not linear (then, nor with chain joins), but it is weakly linear \citep{Meliou2010a}.

\vspace{-2mm}

 \section{CONCLUSIONS}

\vspace{-2mm}
 We have related query causality to abductive diagnosis and the view-update problem. Some connections between the last two
 have been established before. More precisely, the view-update problem has been treated from the point of view of abductive reasoning
\citep{Kakas90,Console95}. The idea is to ``abduce" the presence of tuples in the base tables that explain the presence of those tuples in the view
extension
that one would like, e.g. to get rid of.

In combination with the results reported in \citep{icdt15}, we can see that there are deeper and multiple connections between the areas
of query causality, abductive and consistency-based diagnosis,  view updates, and database repairs. Results for any of these areas can be profitably applied to the others.\footnote{Connections between consistency-based and abductive diagnosis have been established, e.g. in \citep{ConsoleT91}.}

We point out that database repairs are related to the view-update problem.
Actually, {\em answer set programs} (ASPs) \citep{asp} for database repairs \citep{2011Bertossi}  implicity repair the database by updating conjunctive combinations of intentional,
annotated predicates. Those logical combinations -views after all- capture violations of integrity constraints in the original database or along the (implicitly iterative) repair process
(a reason for the use of annotations).

Even more, in \citep{lechen}, in order to protect sensitive information, databases are explicitly and virtually ``repaired" through secrecy views that specify the
information that has to be kept secret. In order to protect
information, a user is allowed to interact only with the virtually repaired versions of the original database that result from making those views empty or
contain only null values. Repairs are specified and computed using ASP, and an explicit connection to prioritized attribute-based repairs \citep{2011Bertossi} is made \citep{lechen}.

Finally, we should note that abduction has also been explicitly applied to database repairs \citep{arieli}. The idea, again, is to ``abduce" possible repair updates that
bring the database to a consistent state.

\vspace{1mm}
\noindent {\bf Acknowledgments:} \ Research funded by NSERC Discovery, and
the NSERC Strategic Network on Business Intelligence (BIN).

{\small

%\begin{thebibliography}{10}

}

\end{document}